# Efficient Kernel based Matched Filter Approach for Segmentation of Retinal Blood Vessels

Sushil Kumar Saroj[a], Vikas Ratna[b], Rakesh Kumar[c], Nagendra Pratap Singh[d]

[a, b, c] Department of Computer Science and Engineering, MMM University of Technology, Gorakhpur, India
[d] Department of Computer Science and Engineering, National Institute of Technology, Hamirpur, India

[a]*skscs@mmmut.ac.in*,[b]*vikascs0113@gmail.com*, [c]*rkiitr@gmail.com*, [d]*nps@nith.ac.in*

**Abstract**

Retinal blood vessel's structure contains information about diseases like obesity, diabetes, hypertension and glaucoma. This information is very useful in identification and treatment of these fatal diseases. To obtain this information, there is need to segment these retinal vessels. Many kernel based methods have been given for segmentation of retinal vessels but their kernels are not appropriate to vessel profile cause poor performance. To overcome this, a new and efficient kernel based matched filter approach has been proposed. The new matched filter is used to generate the matched filter response (MFR) image. We have applied Otsu thresholding method on obtained MFR image to extract the vessels. We have conducted extensive experiments to choose best value of parameters for the proposed matched filter kernel. The proposed approach has examined and validated on two online available DRIVE and STARE datasets. The proposed approach has specificity 98.50%, 98.23% and accuracy 95.77 %, 95.13% for DRIVE and STARE dataset respectively. Obtained results confirm that the proposed method has better performance than others. The reason behind increased performance is due to appropriate proposed kernel which matches retinal blood vessel profile more accurately.

## 1. Introduction

Health is important for human beings to make life better. Many types of diseases are recognized by their symptoms. Some diseases can identify by color and structure. Obesity diabetes, hypertension and glaucoma diseases can be identified by observing retinal blood vessel structure. Glaucoma and diabetes can cause vision loss also. The Ophthalmologist analyzes retinal vessel's structure to identify diseases and diagnose them accordingly. Fundus camera has been used to take image of interior surface of eye. In some retinal images, it is found that the contrast of the retinal vessel is very low compare to its background due to which it is not easy to identify retinal diseases. Therefore, there is need an appropriate method for retinal image segmentation to detect retinal blood vessels accurately. Generally, two kind of methods are used for vessel segmentation. The first method manually segments the blood vessel from an image by the human expert, but the manual processes of segmentation require a lot of time, Practices, Skill and cost. The second method is software-based segmentation. It takes lesser time, efficient and easy to use. Software-based segmentation methods are classified into two broad categories: rule based and machine learning based (Almotri et al.,2018). Both methods have different techniques for segmentation of blood vessels which are shown in Figure 1. In machine learning method, we use supervised learning for blood vessel segmentation. In supervised learning algorithm, trained datasets are utilized to categorize non-vessel and vessel pixels. In the unsupervised algorithm, training datasets are not required. In rule based method, vessel tracking method uses edge based model for example Sobel, Gradient, Prewitt, Robert and Kirsch's differential operator. These edge based methods do not work well in presence of noise and discontinuous nature of retinal vessels. They produce good result only when edges are sharp. Other segmentation approaches are pattern recognition model based mathematical morphology multi scale and kernel based methods.

      Kernel (matched filter) based method is considered best for segmentation of retinal blood vessel compare to other segmentation methods. Matched filter method makes retinal vessels brighter due to which the method identifies vessels more accurately. That is why we choose kernel based method. In the literature, we have designed an efficient kernel by modifying the Gaussian function for better retinal blood vessels segmentation. The major contributions of the work are:

1. Detailed study of Gaussian PDF based kernels design and matched filter approaches for retinal vessel segmentation.

2. Designed the efficient kernel by modifying the Gaussian function for better segmentation of vessels.

3. Conducted the extensive experiments to choose optimal value of parameters for the proposed kernel.

Remaining paper is organized as: related literatures are reviewed in the Section 2, method and proposed model is presented in the Section 3, performance is evaluated in the Section 4. In the last, conclusion of paper with future direction is presented in the Section 5.

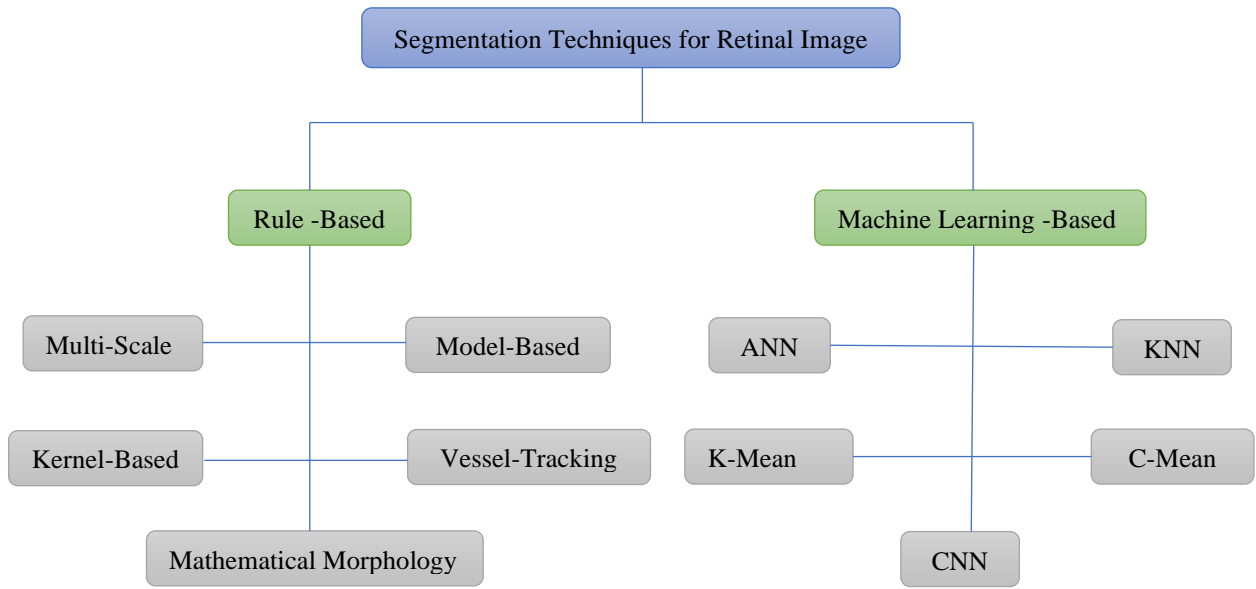

Figure 1: Taxonomy of vessel segmentation techniques

## 2. Literature Review

Segmentation of retinal blood vessel is a critical task. Various approaches are available for segmentation of retinal vessels. Matched filter method is more effective among them. Several matched filter approaches are studied closely and discussed as follows:

Chaudhuri *et al.* (1989): The First 2 Dimensional matched filter approach for retinal blood vessels segmentation was proposed by them. They have used the Gaussian PDF for a kernel design to extract vessels. The author said that the optimal filter has the same shape as an intensity profile of a vessel. In the design of the kernel, they experimentally found that L=9 is extremely good to perform. The limitation of the method is that it considered other bright objects (lesion, optic disk) as a vessel because local contrast of these is high. This degrades performance of the approach.

Al-Rawi *et al.* (2007): They have used the parameters with two-stage matched filter. They have applied the average thresholding by the experiment. The author has optimized the parameters σ and L by conducting experimental tests. Al-Rawi *et al.* have tried to improve the performance by taking search space larger but search space is enough larger.

Singh *et al.* (2015): They have proposed a matched-filter approach based on modified Gaussian PDF and used the entropy-based local thresholding to generate binary image. They have truncated trail of curve and kernel size. The apporach is performed on only DRIVE datasets. The approach is unable to effectively remove retinal boundary and classify them as vessels.

Zhang *et al.* (2010): They have proposed a matched-filter approach i.e. FDOG to identify vessels. Proposed method has two responses; one is matched filter and second is FDOG. The adjusted threshold value is generated by FDOG method. This threshold value is used on the matched filter to segment vessels. This method has used multiple scales to extract the thin and thick vessels. OR logical gate has used to combine them. The problem with this model is that OR logical gate is not much effective to remove the unwanted structure.

Odstrcilik *et al.* (2013): They have proposed a model which has five 2-D kernels. These kernels are designed based on typical cross-sectional profile of retinal vessels. Every filter has different blood vessel width from thinnest to thickest. These kernels are fused with image and maximum response is recorded for every pixel. Problem with the model is that it requires additional computational complexity because of using five kernels and performance is relied on the selection of thick to thin vessels for kernel design.

Chakraborti *et al.* (2015): The author has used a vesselness filter in his model to detect blobs structures applied for self-adaptive matched filter. It is designed from generated output by vesselness filter orientation histogram and relevant information is extracted by the matched filter which present in the orientation histogram. The model overlapped Gaussian curves and require improvement for complex retinal dataset.

Hoover *et al.* (2000): They have given a Gaussian matched-filter and used the probing technique for thresholding. This probe tests an image in a piece and region based properties. Probing method allows multiple regions configuration tested on a pixel. To threshold an image, MFR image histogram is used. The method provides continuous labelling of vessel pixel or non-vessel pixel. The method requires more features.

Kaba *et al.* (2014): The author has performed segmentation task in two stages: pre-processing and probabilistic modeling. Probabilistic model uses joint probability which is distribution function of the Gaussian mixture. To find maximum likelihood estimate of the function expectation, maximization is apply to extract vascular tree. The limitation of model is that it is effective over tracking based scenario only. The model reduces width of vessel compare to original.

Singh *et al.* (2016): has given a new kernel based on Gumbel PDF. They claim that the shape of retinal blood vessels is slightly skewed, and Gumbel has also skewed to its trunk. The actual shape of retinal blood vessels not as Gumbel.

Saroj *et al.* (2020): has given a new kernel based on Frechet PDF. In this paper they claim vessel profile shape similar to Frechet PDF. The design kernel is not appropriate to remove to lesion effectively.

## 3. Proposed Method and Model

The proposed model of retinal blood vessel extraction belongs to class of matched-filter approach. It is the sequence of pre processing, modified Gaussian PDF based efficient kernel design and post processing. The proposed approach has involved various image operations in sequence as follows:

a) **Input RGB colored retinal image**

b) **Pre processing**

   i) Apply PCA method to convert colored retinal image into grayscale image.

   ii) Apply CLAHE method on obtained grayscale image to enhance the quality of grayscale image.

c) **Modified Gaussian PDF based new kernel design**

   i) Take enhanced grayscale image as input.

   ii) Design new matched filter kernel based on modified Gaussian PDF.

   iii) Take kernel size 15 x 17. Generate set of 12 kernels. Rotate them $15^0$ with respect to previous one to cover whole image.

   iv) Convolved the proposed kernel with grayscale image to generate the MFR image.

d) **Post processing**

   i) Apply global Otsu thresholding method to extract the vessels.

   ii) Apply length filtering technique to remove misclassified and isolated pixels.

   iii) Apply masking technique to remove outer artifact.

   iv) Compliment the segmented image for ground-truth comparison.

e) **Finally, the complimented segmented images are used for performance measurement**

### 3.1 Pre processing

Matched filter works on grayscale images. Generally, contrast between retinal blood vessel and its background is low and this contrast gradually decreases as we go outward from optic disk. To generate and enhance grayscale image, pre-processing operation on retinal image is performed. In the paper, pre-processing stage has two parts. In first part, we convert color image into gray-scale image. In second part, quality of output (gray-scale image) of first part is increased.

**3.1.1 RGB to gray scale conversion**

In the proposed approach, we have applied PCA method to convert RGB image into grayscale image because PCA method has low computation complexity and preserve both color and texture discriminability effectively. Gray-scale image produced by PCA is shown in the Figure 2(a). This operation is performed on each image of DRIVE and STARE dataset. This PCA based conversion involves following operations:

1. Red, blue and green color channels are used to vectorized the color image.
2. Compute zero-mean of YCbCr to divide the channels chrominance and luminance.
3. Calculate eigen value of three channels and related normalized eigen vectors.
4. Generate the grayscale image using eigen value weighted linear sum of subspace projection (ECSSP algorithm (Seo *et al.*,2013)).

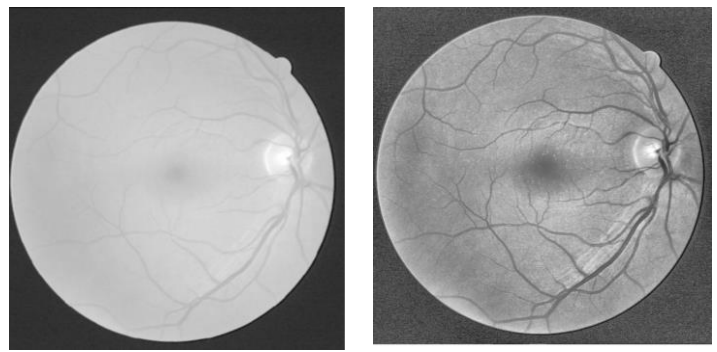

(a): Gray scale image obtained by PCA method     (b): Gray scale image obtained after CLAHE method

Figure 2: Grayscale images generated by various methods for DRIVE

### 3.1.2 Enhancement of grayscale image

We apply CLAHE on grayscale image generated by PCA method to enhance quality of image. The CLAHE method improve the discrimination between vessels and background. Enhanced grayscale images are shown in the Figure 2(b).

### 3.2 Modified Gaussian PDF based new kernel design

In matched filter technique, vessel cross-section profile is compared with pre-design matched filter kernel. In the comparison, their variation and maximum response are stored. In design of matched filter kernel, three points are taken into consideration which are as follows:

1. Generally, blood vessel curvature is small anti-parallel pair and can be detected by piece-wise linear segments.
2. The contrast of vessels with their background is poor and change in intensity profile from vessel to vessel is very low.
3. Width of vessel decreases gradually as we go outward from optic disk.

Chaudhuri *et al.* (1989) have proposed first matched-filter and claimed that vessel cross-section profile is similar to Gaussian curve. Gaussian kernel PDF can be given mathematically as follows:

$$K(x,y) = -e^{(-x^2/2\sigma^2)} \quad \text{for } |y| \leq L/2 \quad (1)$$

where, negative sign shows that contrast of vessel is darker compare to background.

In the literature, we have modified and applied Gaussian function in unique way to design a new efficient kernel. Here to design the proposed kernel, first we have computed value by following equation:

$$K(x,y) = e^{(-x^2/2\sigma^2)} \quad (2)$$

Thereafter, we found max value from $K(x, y)$. We subtracted each value of $K(x, y)$ from the max value. This modification has made in place of using negative sign for inverting the profile. These are represented by an equation as below:

$$K_i'(x,y) = \max(K(x,y)) - K_i(x,y) \quad \text{for } |y| \leq L/2 \quad (3)$$

We subtracted each value of $K(x, y)$ from the max value of $K(x, y)$ because vessel's contrast is low compared to its background. If the background has constant intensity with zero mean additive Gaussian white noise, output of the filter is expected as zero. Therefore, kernel ($K(x, y)$) is modified by subtracting mean value from the function itself which is represented as follows:

$$K_i''(x,y) = K_i'(x,y) - mean(K'(x,y)) \quad \text{for } |y| \leq L/2 \quad (4)$$

Non vessels have higher intensity in comparison to vessels. Therefore, to reduce the detection of non-vessel as vessel, we decrease the intensity of kernel profile by dividing each value of $K''(x, y)$ by total sum value of $K'(x, y)$. This is computed as follows:

$$K_i'''(x,y) = \frac{K_i''(x,y)}{N} \quad (5)$$

where,

$x \rightarrow$ perpendicular distance between point (x, y) and straight line passing through center of the vessel in a direction along its length.

$\sigma \rightarrow$ variance of the intensity profile.

L $\rightarrow$ length of vessel segment which have the same

N $\rightarrow$ sum of all elements in $K'(x,y)$

$i \rightarrow$ value of $i^{th}$ element

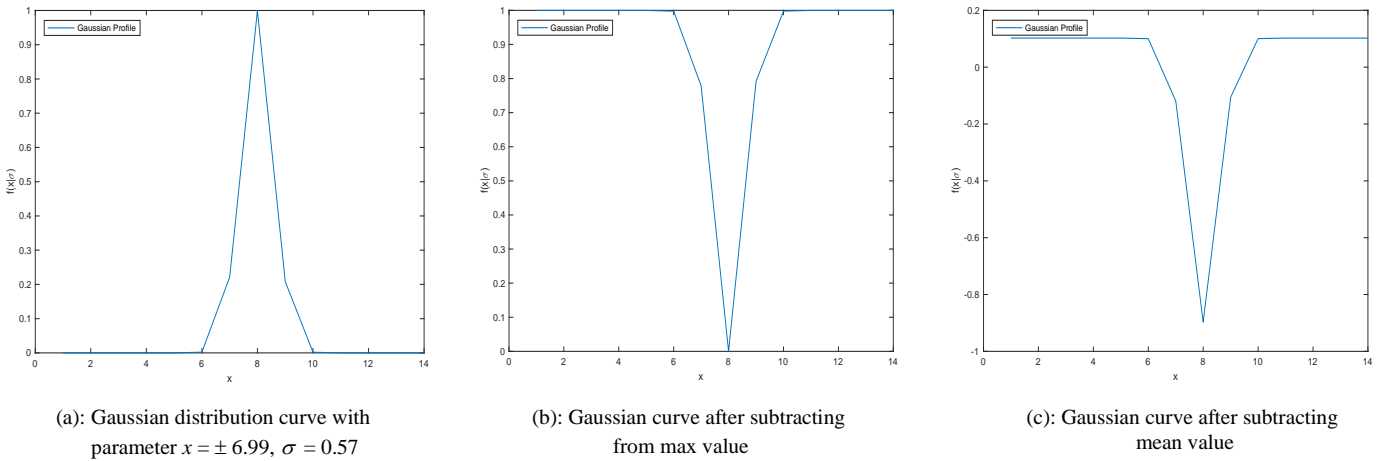

(a): Gaussian distribution curve with parameter $x = \pm 6.99$, $\sigma = 0.57$

(b): Gaussian curve after subtracting from max value

(c): Gaussian curve after subtracting mean value

Figure 3: Operational stages in kernel design for DRIVE dataset

The above operations are performed to design the proposed kernel. Change in curve profile of kernel is shown in the Figure 3. We choose best value of parameters for proposed kernel by conducting exhaustive experimental tests. We choose $x$ from $-6.99$ to $+6.99$. The variance in the scale parameter ($\sigma$) is 0.57 for DRIVE and 1.57 for STARE dataset. Thereafter, we perform the computation for Gaussian PDF by equation 2, Gaussian PDF based kernel by equation 1 and modified Gaussian PDF based proposed kernel by equation 3. The curve profiles obtained from these methods are shown in the Figure 4.

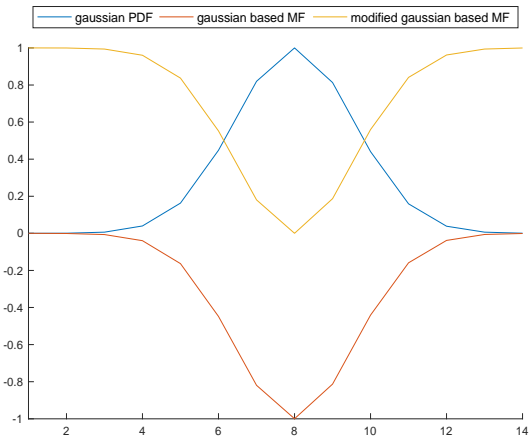

Figure 4: Characteristics profile of Gaussian PDF, Gaussian PDF based kernel and proposed kernel on value $x = \pm 6.99$, $\sigma = 1.57$

We have truncated the tails at $x = \pm 6.99$ because Gaussian curve has infinitely large double sided trails. To choose best value of x, we conducted experimental test for x from value 0.5 to 10 with interval 0.5 and to choose best value of $\sigma$, we conducted experimental test for $\sigma$ from value 0.5 to 10 with interval 0.5 in first round for each image. In the second round, we conducted experimental test on best value of x and $\sigma$ obtained in first round and taking range $\pm 0.5$ around the obtained best value with interval 0.1 for each image. In the third round, we conducted experimental test on best value of x and $\sigma$ obtained in second round and taking range $\pm 0.1$ around the obtained best value with interval 0.01 for each image. Finally, we found $x = \pm 6.99$, $\sigma = 0.57$ at which average accuracy is highest for DRIVE dataset and $x = \pm 6.99$, $\sigma = 1.57$ at which average accuracy is highest for STARE dataset. Graph between different combinations of parameters (x,$\sigma$) and average accuracy for DRIVE datasets are depicted in Figure 5.

In the literature, we determine the length of vessel segment experimentally by examining normal and abnormal retinal vessels. For a non ideal environment, it can reduce the chance of detection of false vessel pixels. We performed experimental test by varying the value of L from 1 to 15 with interval 1. Based on experimental test, it is found that average accuracy increases continuously for value L = 1 to 8 and decreases continuously for value L = 9 to 15 for DRIVE dataset as shown in the Figure 6. In case of STARE dataset, average accuracy increases continuously for value L = 1 to 9 and decreases continuously for value L = 10 to 15. The value L = 8 and L = 9 are selected for DRIVE and STARE datasets respectively.

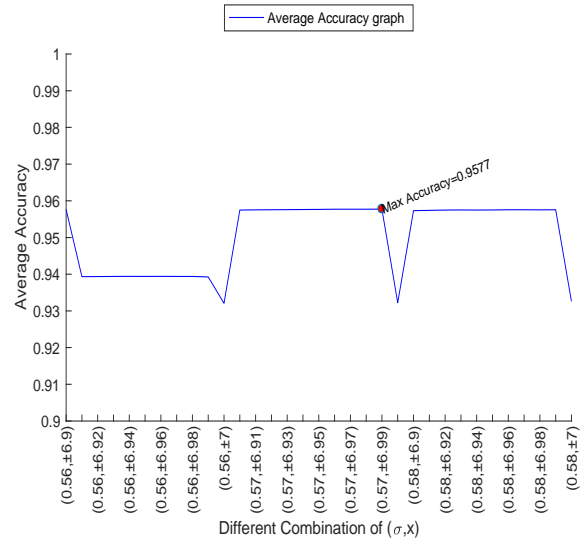

Figure 5: Accuracy of the proposed approach w.r.t distinct combinations of $\sigma$ and $x$ for DRIVE dataset

To design proposed kernel, we use best parameters [x, $\sigma$, L] obtained in the exhaustive experimental tests. In the design of proposed kernel, we suppose a point $P_x = [x, y]$ in the kernel. $\theta_i$ is $i^{th}$ orientation of kernel matched to the vessel profile at angle $\theta_i$ and the kernel is centered about origin [0, 0]. To span piece wise line-segment of an entire retinal image, we require set of kernels.

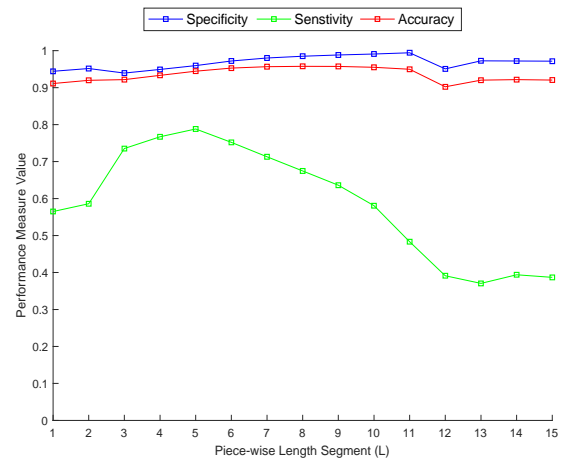

Figure 6: Performance of the proposed approach w.r.t varying length-segment (L) for DRIVE dataset

This set contains 12 kernels and size of kernel is taken 15 x 17. The kernel is rotated at $15^0$ with value of previous state to identify retinal vessels in each direction. We use rotation matrix to calculate 12 kernels by rotating at angle $15^0$. The rotation matrix is given as follows:

$$Rm_i = \begin{bmatrix} \cos\theta_i & -\sin\theta_i \\ \sin\theta_i & \cos\theta_i \end{bmatrix} \quad (6)$$

To cover the whole retinal image, we rotated the kernel with corresponding point $P_{x_i} = [R_x, R_y]$ and rotated coordinate is given by

$$P_{x_i} = [R_x, R_y] = P_x * Rm_i^T \quad (7)$$

These kernels are convolved with grayscale image to generate the MFR image. MFR image produced by proposed matched-filter are shown in the Figure 7.

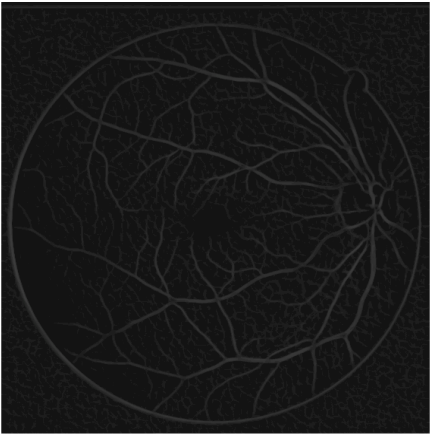

Figure 7: Generated MFR of image '19_test.tif' of DRIVE dataset

**Superiority of proposed kernel over the Gaussian PDF based kernel**

The proposed modified Gaussian PDF based new kernel is efficient and more appropriate to vessel profile in compare to Gaussian PDF based kernel because the proposed kernel inverts the profile in same range value of Gaussian PDF whereas classical Gaussian PDF based kernel inverts the profile in negative range value of Gaussian PDF. We can see these profiles in Figure 4. Therefore, classical Gaussian PDF based kernel profile has higher intensity in comparison to proposed kernel due to which classical Gaussian PDF based kernel detects non vessels, lesions and degenerations as vessels. Due to better matching between proposed kernel and profile of vessel, the kernel classifies vessels more accurately. Therefore, proposed approach has higher accuracy and remove lesions and degenerations efficiently.

**3.3 Post processing**

In post-processing, we performed thresholding, length-filtering, masking and complimenting operations.

**3.3.1 Global Otsu thresholding method**

After matched-filter operation, it is required an effective thresholding algorithm to extract vessel structure from MFR image. Thresholding algorithm generates binary image from MFR image.

We have applied a thresholding technique given by Otsu (1979). This thresholding method uses grayscale image and the image pixel values lies between [0, 255]. Assuming that image pixel is represented in L gray levels i.e. [1, 2,······, L]. $N = n_1 + n_2 + ······+ n_L$ where $n_i$ represents number of pixels present at level of $i$. Histogram of gray level image is normalized and probability distribution is calculated at each intensity level as follows:

$$p_i = \frac{n_i}{N}, \quad p_i \geq 0, \sum_{i=1}^{L} p_i = 1 \quad (8)$$

Thereafter, pixels are divided in to classes $C_o$ and $C_1$ which represent background class for pixel where level is [1, 2, ······, k] and object class for pixel where level is [k+1, k+2,······, L] respectively. Class occurrence probability for background and object are calculated as follows:

$$\omega_o = \sum_{i=1}^{k} p_i \quad (9)$$

$$\omega_1 = \sum_{i=k+1}^{L} p_i \quad (10)$$

Class mean levels for background and object are calculated as follows:

$$\mu_o = \sum_{i=1}^{k} \frac{ip_i}{\omega_o} \quad (11)$$

$$\mu_1 = \sum_{i=k+1}^{L} \frac{ip_i}{\omega_1} \quad (12)$$

Total mean level is computed as follows:

$$\mu_T = \sum_{i=1}^{L} ip_i \quad (13)$$

From above equations, following relations can be verified.

$$\omega_o \mu_o + \omega_1 \mu_1 = \mu_T, \quad \omega_o + \omega_1 = 1 \quad (14)$$

Class variances for background and object are computed as follows:

$$\sigma_o^2 = \sum_{i=1}^{k} \frac{(i - \mu_o)^2 p_i}{\omega_o} \quad (15)$$

$$\sigma_1^2 = \sum_{i=k+1}^{L} \frac{(i - \mu_1)^2 p_i}{\omega_1} \quad (16)$$

For better threshold, we maximize class variance as follows:

$$\sigma_B^2 = \omega_o(\mu_o - \mu_T)^2 + \omega_1(\mu_1 - \mu_T)^2 \qquad (17)$$

Total variance is computed as follows:

$$\sigma_T^2 = \sum_{i=1}^{L}(i - \mu_T)^2 p_i \qquad (18)$$

The optimal threshold $k^*$ that maximizes $\eta$ or equivalently maximizes $\sigma_B^2$ is calculated as follows:

$$\eta(k) = \sigma_B^2(k)/\sigma_T^2 \qquad (19)$$

Finally, the optimal threshold $k^*$ is computed as follows:

$$\sigma_B^2(k) = \max_{1 \leq k < L} \sigma_B^2(k) \qquad (20)$$

Get the desired threshold corresponds to the maximum variance from equation 20. The implemented function ignores the non-zero imaginary part of image. Therefore, the proposed approach uses the Otsu method to segment image. Segmented images (by applying Otsu threshold method) are shown in the Figure 8(a) for images 19_test.tif of DRIVE datasets

### 3.3.2 Length filtering

There may present few misclassified and isolated pixels in segmented image as we can see in the Figure 8(a). We have performed the pixel label propagation and 8-connected neighborhood methods to remove the misclassified and isolate pixels from image. In the Figure 8(b), misclassified and isolate pixels are removed.

### 3.3.3 Masking

There is a chance that some artifacts may exist on retinal borderline. To remove these artifacts, we have applied the concept of masking. Mask images for DRIVE and STARE datasets are available publicly. We take masking of input image and perform the logical AND operation with a segmented image to remove the artifacts. In the Figure 8(c) outer artifacts are removed.

### 3.3.4 Complementing

Finally, we perform complement operation on output image of masking operation. The generated complemented image is analyzed with the ground truth image for performance evaluation. Figure 8(d) show the complemented image.

## 4. Performance Evaluation

### 4.1 Dataset and experimental setup

The proposed approach has been implemented and tested on datasets DRIVE and STARE available online. DRIVE dataset consists 40 images. These 40 images are classified into training and test sets. Both sets contain 20 images each. The Test set contains two sets of manual segmented images while training contains single set of manually segmented images. Mask images for all 40 images are also available showing region of interest. Each image is captured using Canon CR-5 non-mydriatic 3 CCD camera at $45^0$ FOV. These captured images contain 8-bits/plane. These Images are cropped around FOV. Each image has 565 x 584 dimensions. Ground-truth images were segmented manually.

STARE dataset has 400 raw images in which 20 images are selected at random. Out of 20, 10 images are healthy and rest images are unhealthy. These images are captured using TOPCON TRV-50 fundus camera at $35^0$ FOV. These images contain 8 bits/plane. Each image has 700 x 605 dimension. Ground truth images are provided by A. Hoover and V. Kouznetsova.

The proposed method is implemented and tested on MATLAB R2017a with MacBook Pro having configuration 8-GB 1600 MHz RAM, 2.5 GHz dual core i5 processor, Intel HD graphics and 64 bits mac operating system.

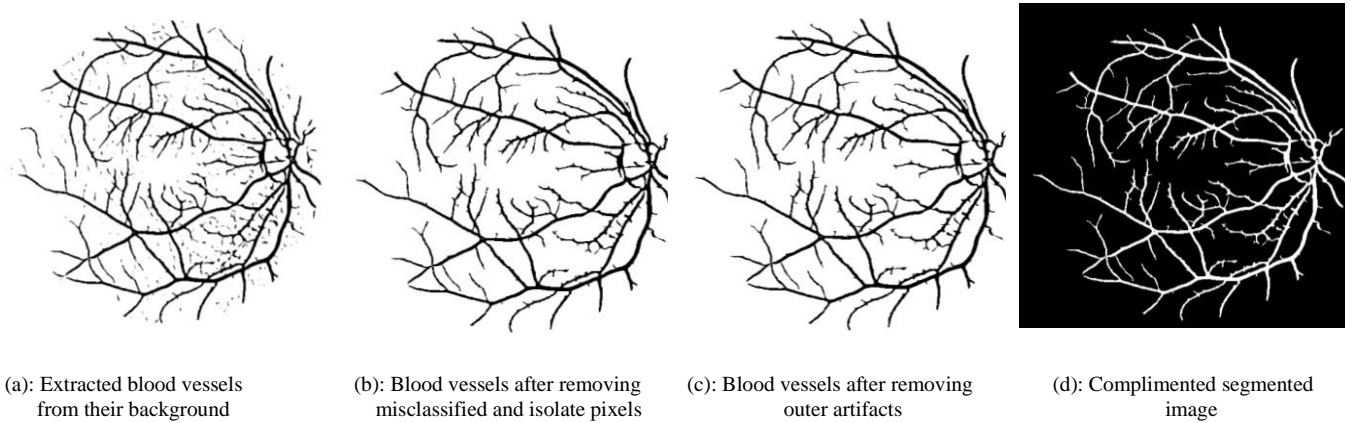

(a): Extracted blood vessels from their background

(b): Blood vessels after removing misclassified and isolate pixels

(c): Blood vessels after removing outer artifacts

(d): Complimented segmented image

Figure 8: Post-processing of image '19_test.tif' of DRIVE dataset

## 4.2 Performance metrics

To compute performance of the method, generated segmented images are compared w.r.t ground truth images. Both images are binary image. Therefore, pixel image has pixel value which is either 1 (true) or 0 (false). When pixel value is true, it treated as a vessel and if false, it treated as non-vessel. On comparing images, there are four outcomes possible which are defined as follows:

1. **True positive (TP):** When segmented image pixel value and respective ground-truth image pixel value is true.
2. **True negative (TN):** When segmented image pixel value and respective ground-truth image pixel value is false.
3. **False positive (FP):** When segmented image pixel value is true while respective ground-truth image pixel value is false.
4. **False negative (FN):** When segmented image pixel value is false while respective ground-truth image pixel value is true.

The table constructed by these four outcomes is known as confusion matrix. On the basis of these four outcomes, we compute the quantitative performance metrics. These performance metrics are applied to compute performance of the method. These performance metrics are described as follows:

**Specificity:** It shows the correctly classified false values from the total number of false values. It is computed as follows:

$$\text{Specificity} = \frac{\text{TN}}{(\text{TN} + \text{FP})} \qquad (21)$$

**Sensitivity:** It shows the correctly classified true values from the total number of true values. It is calculated as follows:

$$\text{Sensitivity} = \frac{\text{TP}}{(\text{TP} + \text{FN})} \qquad (22)$$

**Accuracy:** It shows the correctly classified values from total-number of values. It is calculated as follows:

$$\text{Accuracy} = \frac{\text{TP} + \text{TN}}{(\text{TP} + \text{FN} + \text{TP} + \text{FP})} \qquad (23)$$

In the paper, we also computed the MAD and RMSD quantitative performance metrics.

**RMSD:** It determines the differences between segmented and ground truth images. It shows the deviation of the residuals. If RMSD value is 0 then segmented and ground truth images get superimposed on each other. It is computed as follows:

$$\text{RMSD} = \sqrt{\frac{\sum_{i=1}^{n}(S_i - G_i)}{n}} \qquad (24)$$

where,
  $S_i \rightarrow$ segmented image pixel values.
  $G_i \rightarrow$ ground-truth image pixel values.
  $n \rightarrow$ total number of pixels of an image.

**MAD:** It measures statistical dispersion. It gives difference between actual value and its mean value. If difference between MAD values of segmented and ground truth images is 0, it means both images superimposed on each other. It is computed as follows:

$$\text{MAD} = \sqrt{\frac{\sum_{i=1}^{n} I_i - \bar{I}}{n}} \qquad (25)$$

where,
  $I_i \rightarrow$ pixel value
  $\bar{I} \rightarrow$ mean value
  $n \rightarrow$ total pixels value

**ROC:** To verify effectiveness of segmented image of the proposed method, ROC curve is used. ROC is a 2-D graphical depiction in which x-axis shows false positive rate (FPR) and y-axis shows true positive rate (TPR). AUC is a component of ROC which justifies any segmentation approach effectiveness. AUC values lie between [0, 1]. When value of AUC is 1, it means the vessel and non-vessel are classified 100% correctly. AUC value of a segmentation method is as close as 1 means that segmentation method has better accuracy.

## 4.3 Result analysis

In the literature, various quantitative metrics of performance like sensitivity, specificity and accuracy are calculated to analyse the proposed method. Calculated values of these metrics of the proposed method for all 20 images of DRIVE and STARE datasets are depicted in the Table 1. We have quantitively analysed the proposed method with various existing methods. Values of average specificity, sensitivity and accuracy of proposed approach are compared with other existing methods in the Table 3 and Table 4 for the DRIVE and STARE datasets respectively. The average accuracy of the proposed approach for DRIVE dataset is 0.9577 which is better by 9.06%, 5.02%, 1.23%, 2.04%, 2.47%, 2.12%, 2.16%, 2.97%, 1.74%, 0.57% and 0.34% w.r.t methods given by Chaudhuri *et al.* (1989), Al-Rawi *et al.* (2007), Singh *et al.* (2015), Zhang *et al.* (2010), Odstrcilik *et al.* (2013), Singh *et al.* (2018), Chakraborti *et al.* (2015), Cinsdikici *et al.* (2009), Kaba *et al.* (2014), Singh *et al.*(2016) and Saroj *et al.*(2020) respectively. The average accuracy of the proposed method for STARE dataset is 0.9513 which is better by 2.44%, 4.48%, 1.81%, 1.41%,

0.60%, 2.59%, 6.11%, 2.62% and 0.04% w.r.t approaches introduced by Singh *et al.* (2016), Zhang *et al.* (2010), Odstrcilik *et al.* (2013), Chakraborti *et al.* (2015), Kaba *et al.* (2014), Hoover *et al.* (2014), Singh *et al.* (2018), Singh *et al.*(2016) and Saroj *et al.*(2020) respectively. We also have qualitatively analysed the proposed approach with various existing approaches which can be seen in Figure 9.

We also have computed RMSD, MAD and AUC to validate effectiveness of the proposed method. The proposed method has average RMSD value=0.04, average MAD value=0.03 for DRIVE dataset and average RMSD value=0.05, average MAD value=0.04 for STARE dataset.

Values of RMSD and MAD are close to 0, which means segmented image and ground-truth image is almost superimposed. The RMSD values and difference of MAD values for DRIVE datasets are depicted in the Table 2. TPR (sensitivity) and FPR (1-specificity) are used to plot ROC curve and calculate the AUC from it. The average AUC values are 0.9361 and 0.9221 for DRIVE and STARE datasets respectively. These values are very close to 1, which means the proposed method is efficient for segmentation. ROC curve and AUC for DRIVE datasets are depicted in the Figure 10. Obtained results confirm that the proposed approach has better performance than others.

Table 1: Performance of the method for 20 images of DRIVE & STARE datasets

| DRIVE- Image | Specificity | Sensitivity | Accuracy | STARE- Image | Specificity | Sensitivity | Accuracy |
|---|---|---|---|---|---|---|---|
| 01_test.tif | 0.9782 | 0.7812 | 0.9606 | im0001.ppm | 0.9715 | 0.5100 | 0.9347 |
| 02_test.tif | 0.9876 | 0.7090 | 0.9590 | im0002.ppm | 0.9631 | 0.5565 | 0.9361 |
| 03_test.tif | 0.9676 | 0.7158 | 0.9425 | im0003.ppm | 0.9291 | 0.7432 | 0.9179 |
| 04_test.tif | 0.9904 | 0.6639 | 0.9603 | im0004.ppm | 0.9951 | 0.5126 | 0.9445 |
| 05_test.tif | 0.9906 | 0.6709 | 0.9606 | im0005.ppm | 0.9803 | 0.5344 | 0.9400 |
| 06_test.tif | 0.9905 | 0.6051 | 0.9530 | im0044.ppm | 0.9720 | 0.7545 | 0.9569 |
| 07_test.tif | 0.9875 | 0.6485 | 0.9565 | im0077.ppm | 0.9806 | 0.6636 | 0.9552 |
| 08_test.tif | 0.9895 | 0.5239 | 0.9494 | im0081.ppm | 0.9765 | 0.6693 | 0.9535 |
| 09_test.tif | 0.9950 | 0.5433 | 0.9584 | im0082.ppm | 0.9837 | 0.6550 | 0.9578 |
| 10_test.tif | 0.9853 | 0.6557 | 0.9581 | im0139.ppm | 0.9859 | 0.6682 | 0.9604 |
| 11_test.tif | 0.9842 | 0.6553 | 0.9547 | im0162.ppm | 0.9873 | 0.6162 | 0.9609 |
| 12_test.tif | 0.9873 | 0.6342 | 0.9568 | im0163.ppm | 0.9904 | 0.6911 | 0.9673 |
| 13_test.tif | 0.9872 | 0.6319 | 0.9525 | im0235.ppm | 0.9924 | 0.5668 | 0.9545 |
| 14_test.tif | 0.9804 | 0.7392 | 0.9609 | im0236.ppm | 0.9924 | 0.5587 | 0.9531 |
| 15_test.tif | 0.9690 | 0.7689 | 0.9547 | im0239.ppm | 0.9804 | 0.6772 | 0.9542 |
| 16_test.tif | 0.9867 | 0.7065 | 0.9614 | im0240.ppm | 0.9871 | 0.5206 | 0.9394 |
| 17_test.tif | 0.9905 | 0.5979 | 0.9574 | im0255.ppm | 0.9880 | 0.6574 | 0.9584 |
| 18_test.tif | 0.9866 | 0.7156 | 0.9651 | im0291.ppm | 0.9986 | 0.5929 | 0.9629 |
| 19_test.tif | 0.9779 | 0.8267 | 0.9654 | im0319.ppm | 0.9973 | 0.5867 | 0.9667 |
| 20_test.tif | 0.9881 | 0.6999 | 0.9669 | im0324.ppm | 0.9937 | 0.5633 | 0.9517 |
| **Average** | **0.9850** | **0.6747** | **0.9577** | **Average** | **0.9823** | **0.6149** | **0.9513** |

Table 2: RMSD and MAD values for 20 images of DRIVE dataset

| | im. 1 | im. 2 | im. 3 | im. 4 | im. 5 | im. 6 | im. 7 | im. 8 | im. 9 | im. 10 | img. 11 | im. 12 | im. 13 | im. 14 | im. 15 | im. 16 | im. 17 | im. 18 | im. 19 | im. 20 |
|---|---|---|---|---|---|---|---|---|---|---|---|---|---|---|---|---|---|---|---|---|
| MAD | 0.00 | 0.03 | 0.00 | 0.04 | 0.04 | 0.05 | 0.03 | 0.05 | 0.06 | 0.03 | 0.03 | 0.03 | 0.04 | 0.01 | 0.02 | 0.02 | 0.04 | 0.02 | 0.01 | 0.02 |
| RMSD | 0.04 | 0.04 | 0.06 | 0.04 | 0.04 | 0.05 | 0.04 | 0.05 | 0.04 | 0.04 | 0.05 | 0.04 | 0.05 | 0.04 | 0.05 | 0.04 | 0.04 | 0.03 | 0.03 | 0.03 |

Table 3: Quantitative analysis of the proposed approach with existing approaches for DRIVE dataset

| Author | Segmentation Techniques | Specificity | Sensitivity | Accuracy |
|---|---|---|---|---|
| Chaudhuri *et al.* (1989) | Gaussian M.F | 0.9064 | 0.6326 | 0.8709 |
| Al-Rawi *et al.* (2007) | Improved Gaussian M.F parameters + Automatic thresholding | 0.9553 | 0.5993 | 0.9096 |
| Singh *et al.* (2015) | Modified Gaussian M.F + Local entropy based thresholding | 0.9721 | 0.6735 | 0.9459 |
| Zhang *et al.* (2010) | FDOG M.F + Global thresholding | 0.9724 | 0.7120 | 0.9382 |
| Odstrcilik *et al.* (2013) | Gaussian M.F + Min. error thresholding method | 0.9693 | 0.7060 | 0.9340 |
| Singh *et al.* (2018) | Extended SDGO M.F + Entropy based thresholding | 0.9534 | 0.7738 | 0.9374 |
| Chakraborti *et al.* (2015) | Self adaptive M.F with Gaussian function | 0.9579 | 0.7205 | 0.9370 |
| Cinsdikici *et al.* (2009) | Gaussian M.F and Ant algorithm | - | - | 0.9293 |
| Singh *et al.* (2016) | Gumbel M.F + Entropy based thresholding | 0.9708 | 0.7594 | 0.9522 |
| Saroj *et al.* (2020) | Frechet M.F + Entropy based thresholding | 0.9724 | 0.7278 | 0.9544 |
| Kaba *et al.* (2014) | Gaussian M.F + Probabilistic model using expectation maximization | 0.9683 | 0.7466 | 0.9410 |
| **Proposed Method** | **Modified Gaussian M.F + Otsu thresholding method** | **0.9850** | **0.6747** | **0.9577** |

Table 4: Quantitative analysis of the proposed approach with existing approaches for STARE dataset

| Author | Segmentation Techniques | Specificity | Sensitivity | Accuracy |
|---|---|---|---|---|
| Singh *et al.* (2016) | SDGO M.F + Entropy based thresholding | 0.9423 | 0.7553 | 0.9281 |
| Zhang *et al.* (2010) | FDOG M.F + Global threshold | 0.9736 | 0.7373 | 0.9087 |
| Odstrcilik *et al.* (2013) | Gaussian M.F + Min. error thresholding method | 0.9512 | 0.7847 | 0.9341 |
| Chakraborti *et al.*(2015) | Self adaptive M.F with Gaussian function | 0.9586 | 0.6786 | 0.9379 |
| Kaba *et al.* (2014) | M.F + Probabilistic model using expectation maximization | 0.9672 | 0.7619 | 0.9456 |
| Hoover *et al.* (2000) | Gaussian M.F + Threshold probing | 0.9567 | 0.6751 | 0.9267 |
| Singh *et al.* (2016) | Gumbel M.F + Entropy based thresholding | 0.9376 | 0.7939 | 0.9270 |
| Saroj *et al.* (2020) | Frechet M.F + Entropy based thresholding | 0.9761 | 0.7307 | 0.9509 |
| Singh *et al.* (2018) | Extended SDGO M.F + Entropy based thresholding | 0.9072 | 0.8389 | 0.8931 |
| **Proposed Method** | **Modified Gaussian M.F + Otsu thresholding method** | **0.9823** | **0.6149** | **0.9513** |

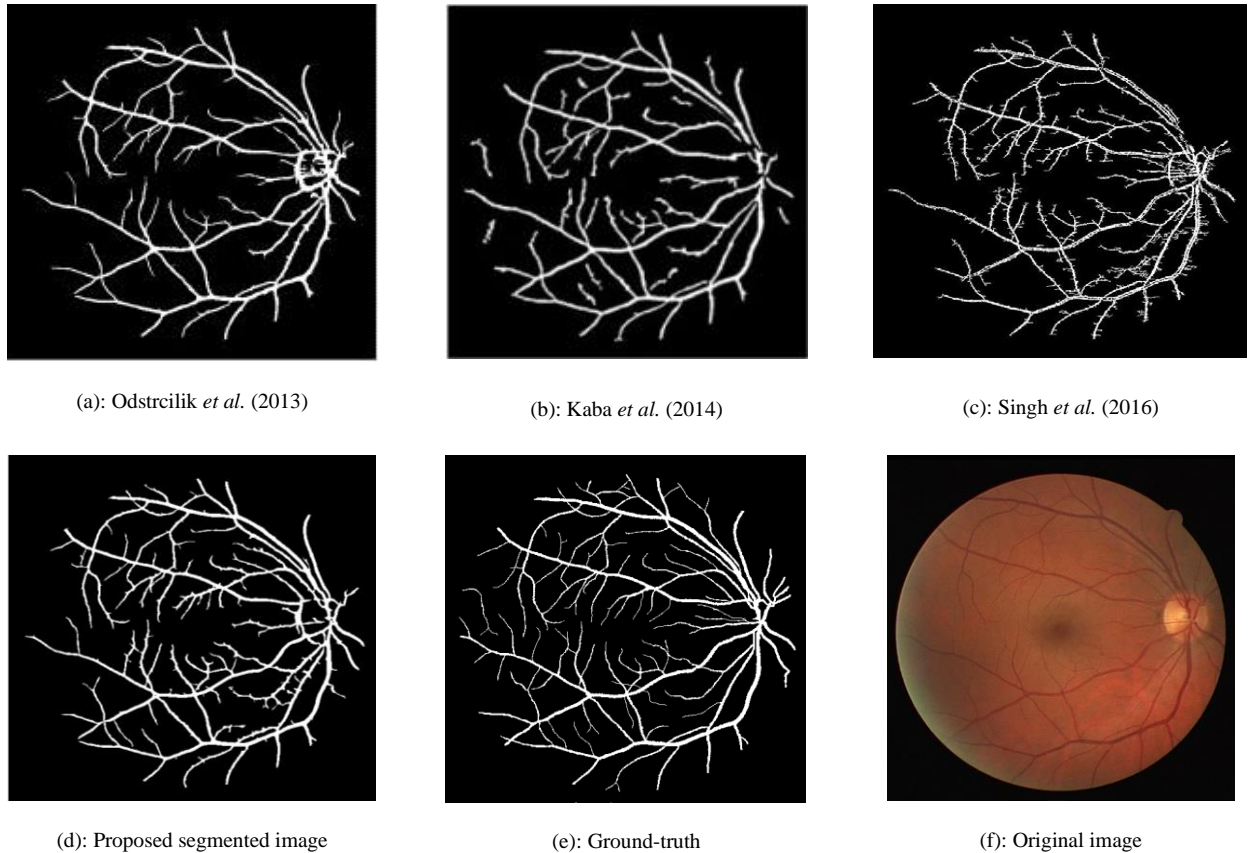

(a): Odstrcilik *et al.* (2013)　　(b): Kaba *et al.* (2014)　　(c): Singh *et al.* (2016)

(d): Proposed segmented image　　(e): Ground-truth　　(f): Original image

Figure 9: Qualitative analysis of the proposed approach with existing methods for image '19_test.tif' of DRIVE dataset

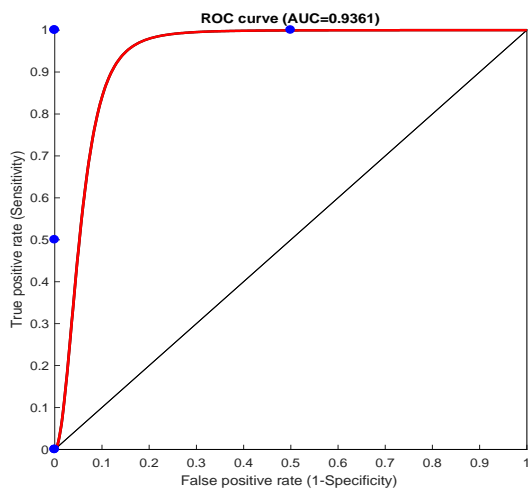

Figure 10: ROC curve for DRIVE dataset

## 5. Conclusion and Future Directions

To diagnose the diseases, it is necessary to identify them first. Information about some diseases such as diabetes, hypertension and glaucoma are present in structure of retinal blood vessels. These diseases are easily identified by segmenting the retinal blood vessels. Several matched-filter methods have been proposed for retinal blood vessels segmentation, but their kernels are not suitable to blood vessel profile causing poor performance. To overcome this, a new and efficient kernel based approach has been proposed. In the literature, we have conducted extensive experiments to opt best value of parameters for the proposed matched filter kernel and use the Otsu thresholding method to finds vessels. The proposed method is validated on two datasets DRIVE and STARE which is available online. Obtained results confirm that the proposed method has better performance than others.

In future work, we will focus on improving the sensitivity by segmenting very thin vessels of retinal image effectively using multi-scale matched filter or others.